# Amino Acids and Their Biological Derivatives Modulate Protein-Protein Interactions In an Additive Way

Xufeng Xu[1*] and Francesco Stellacci[1,2*]

1: Institute of Materials, Ecole Polytechnique Fédérale de Lausanne (EPFL), Lausanne, 1015, Switzerland

2: Bioengineering Institute, Ecole Polytechnique Fédérale de Lausanne (EPFL), Lausanne, 1015, Switzerland

E-mails: xufeng.xu@ru.nl; francesco.stellacci@epfl.ch

**Abstract:**

Protein-protein interactions (PPI) differ when measured in test tubes and cells due to the complexity of the intracellular environment. Free amino acids (AAs) and their derivatives constitute a significant fraction of the intracellular volume and mass. Recently, we have found that AAs have a general property of rendering protein dispersions more stable by reducing the net attractive part of PPI. Here, we study the effects on PPI of different AA derivatives, AA mixtures, and short peptides. We find that all the tested AA derivatives modulate PPI in solution as well as AAs. Furthermore, we show that the modulation effect is additive when AAs form mixtures or are bound into short peptides. Therefore, this study demonstrates the universal effect of a class of small molecules (i.e. AAs and their biological derivatives) on the modulation of PPI and provides insights into rationally designing biocompatible molecules for stabilizing protein interactions and consequently tuning protein functions.

**TOC GRAPHICS**

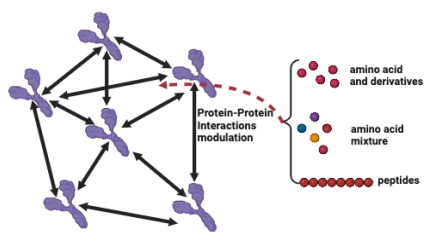

Protein-protein interactions (PPI) are essential for carrying out distinct processes and maintaining homeostasis inside the cell[1]. PPI in cells (*in vivo*) is very hard to reproduce in test tubes (*in vitro*) due to the complexity of intracellular organization[2]. A map of the interactions between proteins and metabolites reveals the importance of intracellular small molecules in modulating protein activity[3,4]. In particular, free amino acids (AAs) were reported to constitute a major component in cellular biomolecules[5]. The total AA concentration in the cytosol of a mammalian cell was also reported to reach around 50 mM in normal conditions[6,7]. The effect of AAs on the protein folding equilibrium between the folded/native and unfolded states (i.e. protein stability) is widely studied (as an important class of osmolytes) by both experimental observation and theoretical descriptions[8,9]. For instance, the protein backbone transfer energies[10] were experimentally measured and a quantitative solvation model was brought up to explain the protecting/denaturing effects of different AAs[11]. $^{19}$F NMR and binding isotherms were also employed[12] to measure the effects of different AAs on the dissociation of a model protein dimer of the B1 domain of the streptococcal immunoglobulin binding protein G and it was found that AAs perturbed the dimer dissociation. Recently, we have found that AAs significantly affect PPI by reducing the net attraction and it in turn makes protein dispersions more stable[13]. The experimental evidence to substantiate this finding was an increase in protein second virial coefficient ($B_{22}$) and a change in protein potential of mean force. We also proposed a theoretical framework of AAs weakly interacting/binding with proteins to explain the AAs' general stabilizing effect. Here, we perform a detailed study of the effects on PPI of different AAs' biological derivatives, of AA mixtures, and of short peptides. We employ lysozyme and bovine serum albumin (BSA) as protein models. They have distinct molecular weights and isoelectric points (14,000 Da and pH 11 for lysozyme; 66,000 Da and pH 4.5 for BSA). We use the method of sedimentation-diffusion equilibrium analytical ultracentrifugation (SE-AUC)[14–17] as an analytical method to measure the second virial coefficient ($B_{22}$)[18,19], which parameter reflects the extent of the solution non-ideality (i.e. protein self-interaction here). We find a general modulation effect on PPI of a variety of AAs' biological derivatives. The modulation effect is related to their aliphatic chain length, hydrophobicity, and charge. We also demonstrate that the modulation effect is additive not only for the mixtures of AAs or of AAs and salt but also for short peptides up to 8 AA residues. The additivity is however lost when peptides are long (1000 ~ 10,000 Da or 40 AA residues). We believe that this study presents a wide class of AA-based molecules capable of stabilizing protein dispersions.

To evaluate protein-protein interactions (PPI) in a dispersion, the equation of state (EOS)[15] (**Equation 1**) was employed:

$$\frac{\Pi}{kT} = \rho_2 + B_{22}\rho_2^2 + B_{23}\rho_2\rho_3 + \cdots \quad (1)$$

where $\Pi$ is the osmotic pressure, $\rho$ is the number density and $kT$ is the product of the Boltzmann constant ($k$) and the temperature ($T$). In the EOS, $B$ indicates the virial coefficient with the number subscripts indicating the component of the dispersion, 1 the solvent, and 2 and 3 the main and minor solute. In this case, $B_{22}$ is the second virial coefficient that measures the self-interaction among the main solutes (i.e. proteins in this study). A positive change of $B_{22}$ ($\Delta B_{22} > 0$) indicates that in the dispersion the net interactions between proteins become more repulsive. In practice, we follow a three-step workflow (**Figure 1**) to measure the effect of small molecules on PPI. 1) In step 1, small molecules at varying concentrations (typically, from 0 up to the solubility limit) are added to a concentrated protein dispersion. 2) In step 2, these as-prepared dispersions are injected into the sample channel of AUC cells by pipettes with the small molecule solution of the same concentration into the reference channel. The AUC cells are then assembled into the AUC rotor. The sedimentation-diffusion equilibrium experiments are performed. After a typical time of approximately 24 h, the sedimentation-diffusion equilibrium is normally achieved and the raw data detailing the protein concentration gradient along the AUC cell radius are collected. 3) In step 3, we perform the data analysis. It involves: i) calculating osmotic pressure by integrating protein concentration along the radius; ii) computing the $B_{22}$ values by assessing the slope of osmotic pressure divided by protein concentration as a function of protein concentration and iii) generating a plot of the $B_{22}$ change ($\Delta B_{22}$) as a function of small molecule concentration (detailed calculation steps and related theoretical equations in **Methods**). A typical example of data analysis is shown in the red box in **Figure 1**. The $\frac{\Pi}{kT}$ vs $\rho$ curve is above the van't Hoff line (dashed), which indicates that the net lysozyme-lysozyme interactions in dispersion are repulsive ($B_{22} > 0$)[15]. Then, the value for $B_{22}$ can be obtained by linear fitting (red dashed) the curve of $\frac{\Pi}{kT\rho}$ vs $\rho$: $1.08 \times 10^{-25}\ m^3$. The data in the small $\rho$ region is omitted from the linear fitting due to data noise in low protein concentrations. Using this workflow, the effect of any small molecules on PPI can be measured. As shown in the blue box in **Figure 1**, the effects of different AAs on lysozyme-lysozyme interactions are obtained by this workflow in our previous study[13]. It is noteworthy that the molecular mass of small molecules has to be significantly smaller than that of proteins to ensure negligible small molecule sedimentation during the sedimentation process of the proteins.

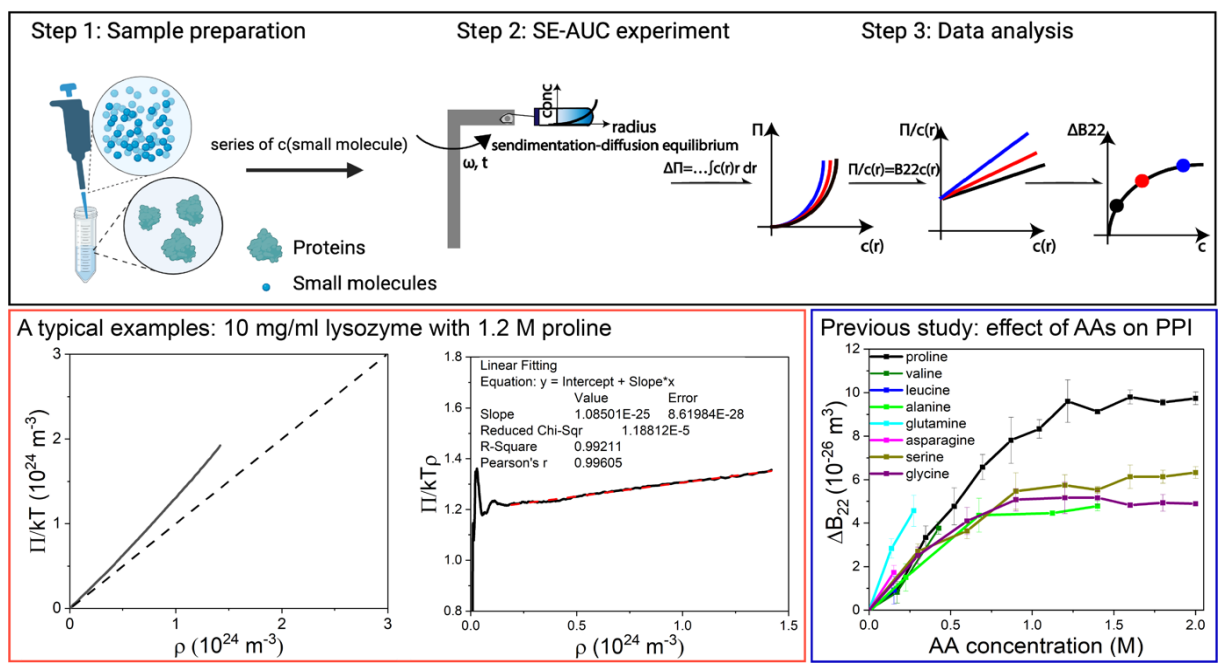

**Figure 1: Workflow to measure the effect of small molecules on protein-protein interactions.** Representative scheme of the workflow (black box) to measure the effect of small molecules on protein-protein interactions (characterized by $B_{22}$). It consists of three steps: 1. Sample preparation; 2. sedimentation-diffusion equilibrium analytical ultracentrifugation (SE-AUC) experiment and 3. Data analysis. Created by BioRender.com; A typical example of measuring $B_{22}$ for lysozyme-lysozyme interactions in 10 mg/ml lysozyme dispersion with 1.2 M proline (red box) and a previous study[13] of the effects of different AAs on lysozyme-lysozyme interactions by using this workflow (blue box).

**Amino acid derivatives:** We firstly studied the effect of chirality. As shown in **Figure 2A**, we found that the modulation of the two proline enantiomers (D- and L-proline) on $\Delta B_{22}$ for lysozyme-lysozyme interactions basically overlaps. We also tested a 1:1 racemic mixture of D- and L-proline (**Figure S1**), and the effect is the same as a pure enantiomer (L-proline). Then, we investigated the effect of amine group location on the AA. We employed alpha-alanine, beta-alanine, and gamma-aminobutyric acid (GABA), where the amine group is attached to the alpha-carbon, beta-carbon and gamma-carbon respectively. As shown in **Figure 2B**, we found that $\Delta B_{22}$ does not change when the amine group is switched from the alpha- to beta-carbon. However, changing the amine group location to gamma-carbon (with the addition of one methyl group) increased $\Delta B_{22}$. This indicates that the effect of AAs on $\Delta B_{22}$ may be not dependent on amine group location but on aliphatic chain length. Based on this hypothesis, a further study was conducted by using 3 diols from 1,2-ethylene glycol (2-carbon aliphatic chain) to 1,4-butanediol (4-carbon aliphatic chain) and 1,6-hexanediol (6-carbon aliphatic chain) which only differ in aliphatic chain lengths. As shown in **Figure S2**, we found that the effect of diols on $\Delta B_{22}$ increases gradually with longer aliphatic chain lengths and 1,6-hexanediol has the best stabilization effect on the protein solution, which could explain the widespread use of 1,6-hexanediol in biological assays to dissolve protein phase separation[20]. A diol of a longer aliphatic chain length than 7 carbons could not be tested due to its limited water solubility. We also studied the effect of glucose, a main product of the catabolism of AAs[21] and we found a weaker effect on $\Delta B_{22}$ (**Figure 2C**), which may be explained by a lower hydrophobicity of glucose compared to AAs.

Proteins normally change their properties by post-translational modifications (PTMs)[22], where different functional groups are covalently bonded to AAs. Three main PTMs of AAs were studied here. As shown in **Figure 2D**, we compared the effects of proline and hydroxyl-proline (the most frequently hydroxylated AA residue in the human body[23]). We found that the hydroxylation decreases the effect of proline on $\Delta B_{22}$, which may be due to lower hydrophobicity after the hydroxylation modification[23]. Both the phosphorylation and acetylation of AAs were also investigated as they introduce negative charge to AAs. We found that the both PTMs enhance the stabilization effect on lysozyme-lysozyme interactions (**Figure 2D**). This may be due to more negative charge after the PTMs, enhancing the binding of these AAs on positively charged lysozyme[13]. The destabilization effect was found when negatively charged BSA was employed since the more negative charge by the phosphorylation and acetylation, the less binding to negatively charged BSA (-) (**Figure 2D**). The same phenomenon was observed for the effects of negatively charged arginine (-) on lysozyme (+) and BSA (-) (**Figure S3**).

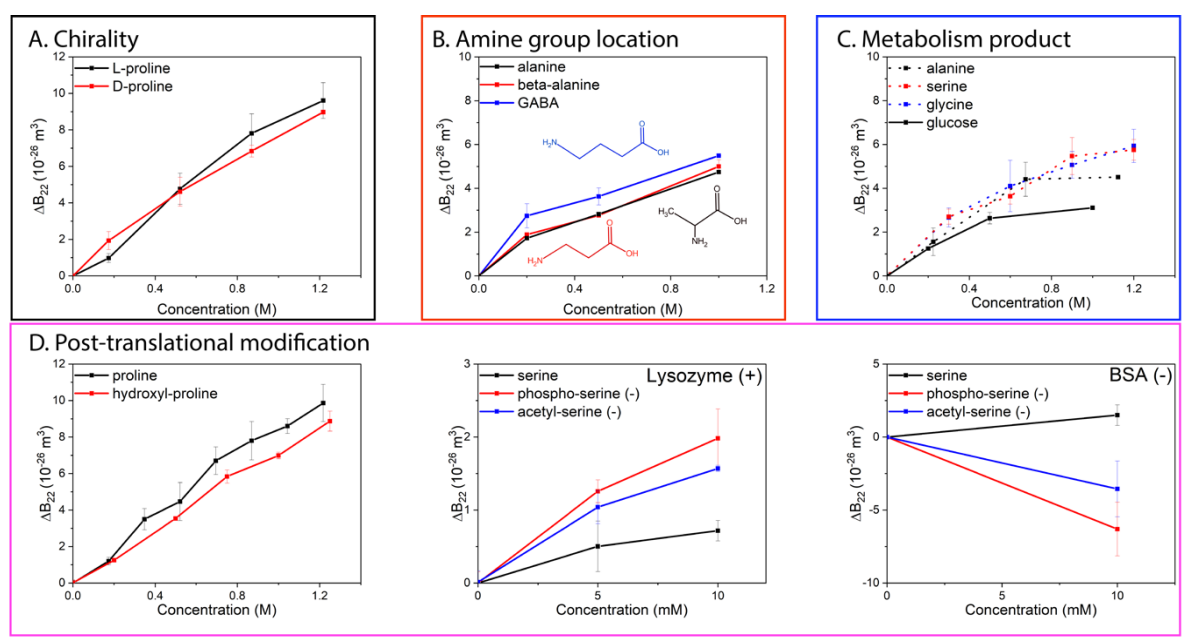

**Figure 1**: **The effect on protein-protein interactions of different AA derivatives. Plots of $\Delta B_{22}$ vs concentration of small molecules.** A. *Chirality*: the effect of two proline enantiomers (D- and L-proline) on lysozyme-lysozyme interactions; B. *Amine group location*: the effect of alanine, beta- alanine, and GABA on lysozyme-lysozyme interactions; C. *Metabolism product*: the effect of glucose on lysozyme-lysozyme interactions, compared with three AAs (alanine, serine and glycine); D. *Post-translational modification*: the effect of proline and hydroxyl-proline on lysozyme-lysozyme interactions; the effect of serine, phospho-serine, and acetyl-serine on lysozyme-lysozyme interactions and BSA-BSA interactions.

**Amino acid mixtures:** As shown in **Figure 3A**, we found that the effect of a binary mixture of glycine (0.6 M) and proline (0.5 M) equals to the addition of the separate effects from the two AAs. A quinary AA mixture was also employed. It consisted of glycine, alanine, asparagine, proline, and serine 0.1 M each (the same composition as in Gibco™ MEM Non-Essential Amino Acids Solution for cell culture media). We found that the effect of this complex AA mixture at a total AA concentration of 0.5 M equals to the effect of proline at a concentration of 0.5 M (**Figure 3B**). In a previous report, we showed that when adding proline to a solution that contained NaCl we could counter the salt de-stabilization with the stabilization effect of proline. Here we revisit the same system and show that the salt and proline effects are indeed simply additive. As shown in **Figure 3C** and **3D**, we mixed 1 M proline with 0.2 M or 0.3 M NaCl respectively. We found that the overall effect of the mixture is roughly the addition of the effects from proline and NaCl. Therefore, the modulation effect of AAs is shown to be additive when different AAs are mixed or AAs are mixed with salt.

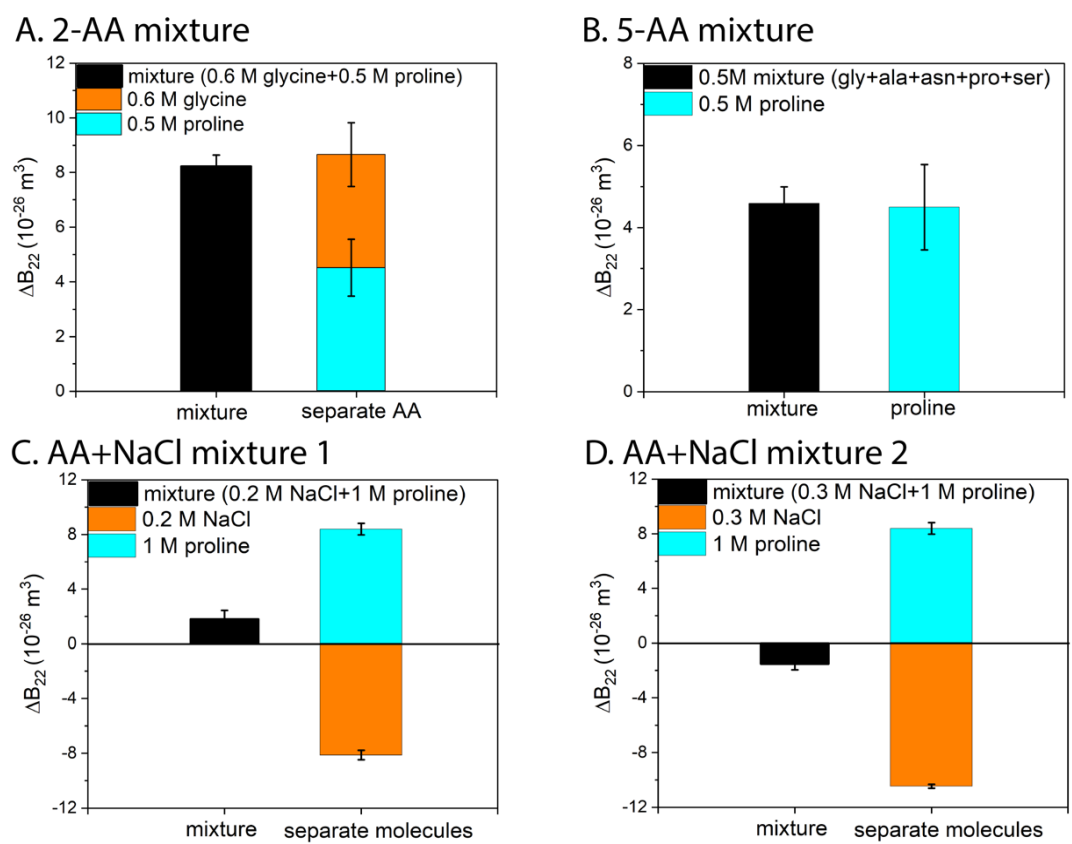

**Figure 3**: **The effect on protein-protein interactions of different AA mixtures. Plots of $\Delta B_{22}$ vs mixture concentration.** The effect on lysozyme-lysozyme interactions of A. 0.6 M glycine, 0.5 M proline, and the mixture; B. a complex AA mixture used in MEM Non-Essential Amino Acids Solution, consisting of glycine, alanine, asparagine, proline, and serine of 0.1 M, and proline of 0.5 M; C. 0.2 M NaCl, 1 M proline, and the mixture; D. 0.3 M NaCl, 1 M proline, and the mixture.

**Peptides:** In our previous publication[13] we showed that peptide made of three of four proline residues have a roughly additive effect when compared to proline. Here we investigated further this effect. First, we show in **Figure 4A**, that the effect of a poly(proline) being additive in not only true in (proline)$_3$, (proline)$_4$ but also in (proline)$_8$. However, the depletion interactions[24] take over when the molecular weight for the poly(proline) increases to 1000 ~ 10,000 Da. $\Delta B_{22}$ becomes negative when poly(proline) was added into BSA solution, indicating more attractive PPI, in stark contrast to positive $\Delta B_{22}$ (repulsive PPI) for proline at the same mass concentration (**Figure 4B**). We also investigated the effect of hetero dipeptides made of proline and glycine, including H-pro-gly-OH, H-gly-pro-OH, and cyclo-(pro-gly). As shown in **Figure 4C**, the effects of these three dipeptides are similar to the additive effects of proline and glycine. However, when the peptide length is increased to 40 amino acid residues the depletion interactions again took over. As shown in **Figure 4D**, $\Delta B_{22}$ becomes negative for poly(pro-gly)$_{20}$ compared to positive $\Delta B_{22}$ for proline. These two experiments on polyAA and (poly)dipeptides indicates that the modulation effect is still additive when AAs are bound into short peptides. However, the additivity does not hold when the peptide becomes long due to depletion attraction.

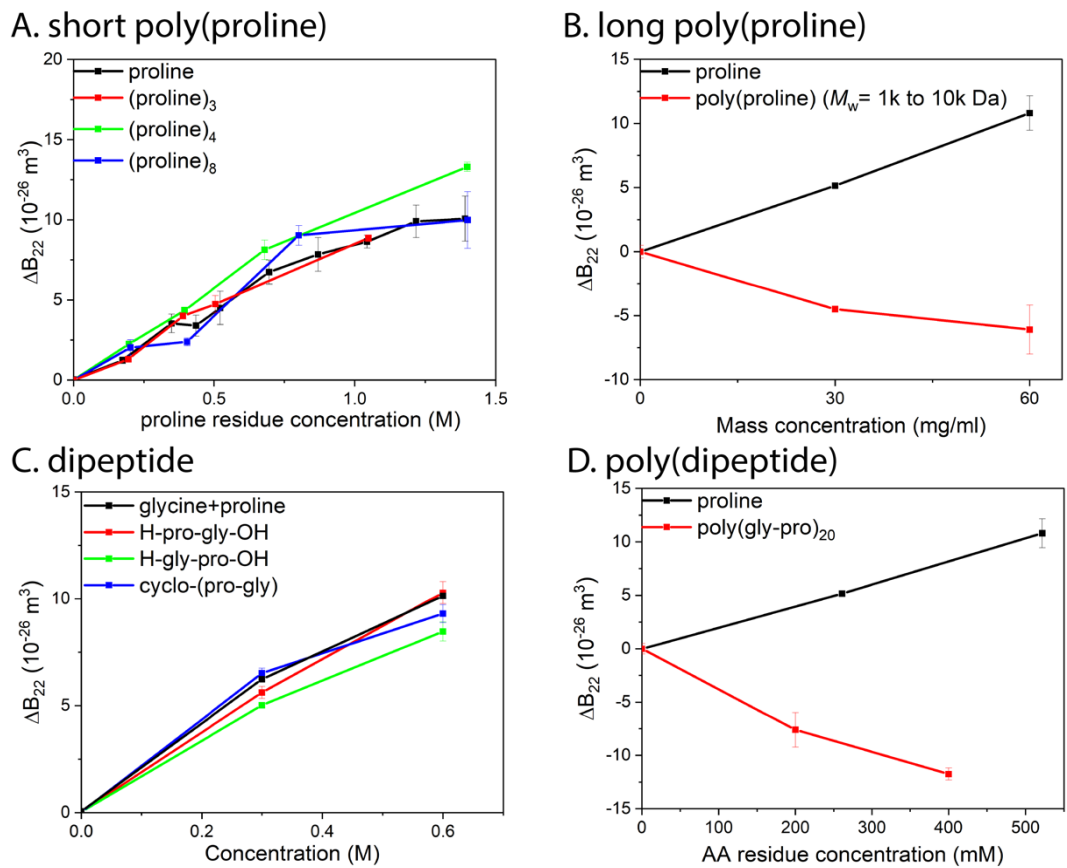

**Figure 4**: **The effect on protein-protein interactions of different peptides. Plots of $\Delta B_{22}$ vs peptide concentration.** A. The effect on lysozyme-lysozyme interactions of three homopeptides of proline (including (proline)$_3$, (proline)$_4$ and (proline)$_8$); B. The effect on BSA-BSA interactions of proline and of the polypeptide of proline, poly(proline) with $M_w$ = 1000 ~ 10,000 Da; C. The effect on lysozyme-lysozyme interactions of the three dipeptides made of proline and glycine, including H-pro-gly-OH, H-gly-pro-OH, and cyclo-(pro-gly) compared to the additive effects of proline and glycine; D. The effect on BSA-BSA interactions of proline and of the polypeptide made of glycine and proline poly(gly-pro)$_{20}$. BSA was chosen for studying the effects of poly(proline)

(1000 ~ 10,000 Da) and poly(gly-pro)$_{20}$ due to the significantly larger molecular weight of BSA (66,000 Da) than that of peptides.

**Solution pH**: In this study, we have consistently found that when the addition of small molecules changes the solution pH, the $\Delta B_{22}$ measured reflects the change of electrostatic repulsion between proteins rather than the effect of small molecules on PPI. For instance, we found that the effect of a heptapeptide (H-Ser-Leu-Ser-Leu-Ser-Pro-Gly-OH) is more significant than the additive effects of all the AA residues and the difference is more pronounced at higher concentrations (**Figure S4A**). This discrepancy is due to the buffer pH change after the addition of the peptide. As shown in **Figure S4B**, the buffer pH becomes lower with the addition of the peptide, which makes the lysozyme more electrostatically repulsive. This explains the significantly larger $\Delta B_{22}$ value for the peptide compared to the addition of the effects of all AA residues. Therefore, it is crucial to ensure an unchanged buffer pH after the addition of any small molecules for the investigation of any potential effect of small molecules on PPI.

In this study, all the biological derivatives of AAs were shown to modulate PPI. The change in the AA chirality and the amine group location does not affect AAs' modulation on PPI. Instead, the aliphatic chain length of AAs is found to affect their modulation on PPI, which is further illustrated by using a series of diols of 2, 4, and 6 carbon chain lengths. The PTMs including phosphorylation and acetylation introduce more charge to AAs and affect their modulation on PPI: the opposite charge between AAs and proteins help AAs stabilize the PPI while the same charge destabilizes PPI, which can be explained by our recently proposed theory of AAs weakly binding on protein[13]. Moreover, we show that AAs have an additive effect when they form mixtures or short peptides. However, long peptides destabilize PPI by introducing depletion attraction. Special cautions should be taken for plausible buffer pH change after the addition of small molecules as the buffer pH change will significantly affect electrostatic repulsion between proteins. Overall, this study demonstrated the general effects of a wide class of small molecules (i.e. AAs and their biological derivatives) on PPI. The modulation effect can be also extended to AA mixtures and short peptides by the simple additivity principle. Therefore, this knowledge improves our understanding of the structure principles of AA-related molecules for the modulation on PPI, which can be used to rationally design AA-based small molecules for controlling protein stability and consequently function.

## Materials and Methods:

### Materials

The powders of bovine serum albumin (BSA) and amino acids (including L-proline, L-alanine, L-serine L-glycine, D-proline, beta-alanine, gamma-aminobutyric acid (GABA), hydroxyl-proline, arginine-HCl, phospho-serine, and acetyl-serine) were purchased from Sigma-Aldrich. 1,2-ethylene glycol, 1,4-butanediol, and 1,6-hexanediol were purchased from Fisher Scientific. The powder of lysozyme was purchased from Carl Roth. All the peptides (including (pro)$_3$, (pro)$_4$, (pro)$_8$, H-pro-gly-OH, H-gly-pro-OH, cyclo-(pro-gly), and H-Ser-Leu-Ser-Leu-Ser-Pro-Gly-OH) were purchased from Bachem. Poly(proline) ($M_w$: 1,000-10,000) was purchased from Sigma-Aldrich. Poly(gly-pro)$_{20}$ was purchased from GenScript.

### Methods

Sedimentation-Diffusion Equilibrium:

In a typical SE experiment[15,25], an analytical ultracentrifuge (Beckman Coulter ProteomeLab XL-I/XL-A) and titanium double sector cells of 3 mm pathlength were used. Protein solutions (e.g. 10 mg/ml) with different concentrations of AAs in 50 mM phosphate buffer (pH 7.1) were prepared. The solution of an appropriate volume (e.g., 60 μL) was added to the sample channel and the AA solution of the same concentration and volume in the buffer was added to the reference channel in an AUC cell by a pipette. SE experiments were then performed overnight at 20 °C with scan intervals of 2 h using interference and absorbance optics (radial steps: 3 μm) at an appropriate angular velocity (44,000 rpm for lysozyme and 24,000 rpm for BSA). Typically, a sedimentation-diffusion equilibrium was reached after 24 h, when the concentration profile stayed unchanged for at least 6 h.

Data analysis:

The raw data from a SE-AUC experiment is the fringe difference ($\Delta J$) versus radial positions ($r$). $\Delta J$ was firstly converted to concentration difference ($\Delta c$) by using **Equation 2**[25,26], where λ= 655 nm for the laser diode light source embedded in the Optima XLI, d$n$/d$c$ is the refractive index increment, which the abbe refractometer can measure. $a$ is the pathlength of the AUC cell, which equals 3 mm.

$$\Delta c = \frac{\Delta J \lambda}{\frac{dn}{dc}a} \quad (2)$$

The concentration gradient ($\Delta c$ versus $r$) was then converted to an equation of state curve (osmotic pressure $\Pi$ versus protein number density $\rho$) by using **Equation 3**[27,28] where $\omega$ the angular velocity, $\Delta m$ the protein buoyant mass.

$$\Delta \Pi = \omega^2 \Delta m \int_{r_m}^{r_1} \rho(r)\, r\, \mathrm{d}r \quad (3)$$

$B_{22}$ was calculated by the equation of state (EOS) (**Equation 1**).

The final step was to calculate $\frac{\Pi}{\rho_2}$ and the slope for $\frac{\Pi}{\rho_2}$ versus $\rho_2$ equals $B_{22}$[15] (**Equation 4**).

$$\frac{\Pi}{kT\rho_2} = (1 + B_{23}\rho_3) + B_{22}\rho_2 \quad (4)$$

The $B_{22}$ variation ($\Delta B_{22}$) was calculated by using **Equation 5** where $B_{22}^0$ the value of $B_{22}$ without any AA addition.

$$\Delta B_{22} = B_{22} - B_{22}^0 \quad (5)$$

**Supporting Information:**

Supporting Information is available.

**Notes:**

The authors declare no competing financial interests.

**Acknowledgements:**

X.X. and F.S. acknowledge the support of the European Union's Horizon 2020 Research and Innovation program under grant agreement no. 101017821 (LIGHT-CAP).

**References:**

(1) Hein, M. Y.; Hubner, N. C.; Poser, I.; Cox, J.; Nagaraj, N.; Toyoda, Y.; Gak, I. A.; Weisswange, I.; Mansfeld, J.; Buchholz, F.; Hyman, A. A.; Mann, M. A Human Interactome in Three Quantitative Dimensions Organized by Stoichiometries and Abundances. *Cell* **2015**, *163* (3), 712–723. https://doi.org/10.1016/j.cell.2015.09.053.
(2) Cohen, R. D.; Pielak, G. J. A Cell Is More than the Sum of Its (Dilute) Parts: A Brief History of Quinary Structure. *Protein Science* **2017**, *26* (3), 403–413. https://doi.org/10.1002/pro.3092.
(3) Piazza, I.; Kochanowski, K.; Cappelletti, V.; Fuhrer, T.; Noor, E.; Sauer, U.; Picotti, P. A Map of Protein-Metabolite Interactions Reveals Principles of Chemical Communication. *Cell* **2018**, *172* (1), 358-372.e23. https://doi.org/10.1016/j.cell.2017.12.006.
(4) Gerosa, L.; Sauer, U. Regulation and Control of Metabolic Fluxes in Microbes. *Current Opinion in Biotechnology* **2011**, *22* (4), 566–575. https://doi.org/10.1016/j.copbio.2011.04.016.
(5) Dar, F.; Pappu, R. Restricting the Sizes of Condensates. *eLife* **2020**, *9*, e59663. https://doi.org/10.7554/eLife.59663.
(6) Gauthier-Coles, G.; Vennitti, J.; Zhang, Z.; Comb, W. C.; Xing, S.; Javed, K.; Bröer, A.; Bröer, S. Quantitative Modelling of Amino Acid Transport and Homeostasis in Mammalian Cells. *Nat Commun* **2021**, *12* (1), 5282. https://doi.org/10.1038/s41467-021-25563-x.
(7) Bergström, J.; Fürst, P.; Norée, L. O.; Vinnars, E. Intracellular Free Amino Acid Concentration in Human Muscle Tissue. *Journal of Applied Physiology* **1974**, *36* (6), 693–697. https://doi.org/10.1152/jappl.1974.36.6.693.
(8) Rose, G. D.; Fleming, P. J.; Banavar, J. R.; Maritan, A. A Backbone-Based Theory of Protein Folding. *Proceedings of the National Academy of Sciences* **2006**, *103* (45), 16623–16633. https://doi.org/10.1073/pnas.0606843103.

(9) Hu, C.; Rösgen, J.; Pettitt, B. M. Osmolyte Influence on Protein Stability: Perspectives of Theory and Experiment. In *Modeling Solvent Environments*; John Wiley & Sons, Ltd, 2010; pp 77–92. https://doi.org/10.1002/9783527629251.ch4.

(10) Auton, M.; Bolen, D. W. Additive Transfer Free Energies of the Peptide Backbone Unit That Are Independent of the Model Compound and the Choice of Concentration Scale. *Biochemistry* **2004**, *43* (5), 1329–1342. https://doi.org/10.1021/bi035908r.

(11) Street, T. O.; Bolen, D. W.; Rose, G. D. A Molecular Mechanism for Osmolyte-Induced Protein Stability. *Proceedings of the National Academy of Sciences* **2006**, *103* (38), 13997–14002. https://doi.org/10.1073/pnas.0606236103.

(12) Rydeen, A. E.; Brustad, E. M.; Pielak, G. J. Osmolytes and Protein–Protein Interactions. *J. Am. Chem. Soc.* **2018**, *140* (24), 7441–7444. https://doi.org/10.1021/jacs.8b03903.

(13) Xufeng Xu, Ting Mao, Pamina M. Winkler, Francesco Stellacci. Amino Acids Effect on Colloidal Properties [Unpublished Manuscript].

(14) Schuck, P.; Braswell, E. H. Measuring Protein-Protein Interactions by Equilibrium Sedimentation. *Current Protocols in Immunology* **2000**, *40* (1), 18.8.1-18.8.22. https://doi.org/10.1002/0471142735.im1808s40.

(15) Xu, X.; Ong, Q.; Mao, T.; Silva, P. J.; Shimizu, S.; Rebecchi, L.; Kriegel, I.; Stellacci, F. Experimental Method to Distinguish between a Solution and a Suspension. *Advanced Materials Interfaces* **2022**, *9* (19), 2200600.

(16) Xu, X.; With, G. de; Cölfen, H. Self-Association and Gel Formation during Sedimentation of like-Charged Colloids. *Materials Horizons* **2022**, *9* (4), 1216–1221. https://doi.org/10.1039/D1MH01854J.

(17) Xu, X.; Cölfen, H. Ultracentrifugation Techniques for the Ordering of Nanoparticles. *Nanomaterials* **2021**, *11* (2), 333.

(18) Wills, P. R.; Scott, D. J.; Winzor, D. J. The Osmotic Second Virial Coefficient for Protein Self-Interaction: Use and Misuse to Describe Thermodynamic Nonideality. *Anal Biochem* **2015**, *490*, 55–65. https://doi.org/10.1016/j.ab.2015.08.020.

(19) Le Brun, V.; Friess, W.; Schultz-Fademrecht, T.; Muehlau, S.; Garidel, P. Lysozyme-Lysozyme Self-Interactions as Assessed by the Osmotic Second Virial Coefficient: Impact for Physical Protein Stabilization. *Biotechnology Journal* **2009**, *4* (9), 1305–1319. https://doi.org/10.1002/biot.200800274.

(20) Kroschwald, S.; Maharana, S.; Simon, A. Hexanediol: A Chemical Probe to Investigate the Material Properties of Membrane-Less Compartments. *Matters* **2017**. https://doi.org/10.19185/matters.201702000010.

(21) Newsholme, P.; Stenson, L.; Sulvucci, M.; Sumayao, R.; Krause, M. 1.02 - Amino Acid Metabolism. In *Comprehensive Biotechnology (Second Edition)*; Moo-Young, M., Ed.; Academic Press: Burlington, 2011; pp 3–14. https://doi.org/10.1016/B978-0-08-088504-9.00002-7.

(22) Tak, I.-R.; Ali, F.; Dar, J. S.; Magray, A. R.; Ganai, B. A.; Chishti, M. Z. Chapter 1 - Posttranslational Modifications of Proteins and Their Role in Biological Processes and Associated Diseases. In *Protein Modificomics*; Dar, T. A., Singh, L. R., Eds.; Academic Press, 2019; pp 1–35. https://doi.org/10.1016/B978-0-12-811913-6.00001-1.

(23) Hydroxylation of Proteins. In *Co‐ and Post-Translational Modifications of Therapeutic Antibodies and Proteins*; John Wiley & Sons, Ltd, 2019; pp 119–131. https://doi.org/10.1002/9781119053354.ch10.

(24) Tuinier, R. Introduction to Depletion Interaction and Colloidal Phase Behaviour. In *Soft Matter at Aqueous Interfaces*; Lang, P., Liu, Y., Eds.; Springer International Publishing: Cham, 2016; pp 71–106. https://doi.org/10.1007/978-3-319-24502-7_3.

(25) Xu, X.; Franke, T.; Schilling, K.; Sommerdijk, N. A. J. M.; Cölfen, H. Binary Colloidal Nanoparticle Concentration Gradients in a Centrifugal Field at High Concentration. *Nano Letters* **2019**, *19* (2), 1136–1142. https://doi.org/10.1021/acs.nanolett.8b04496.

(26) Maechtle, W.; Börger, L. *Analytical Ultracentrifugation of Polymers and Nanoparticles*; Springer laboratory; Springer: Berlin ; New York, 2006.

(27) Page, M. G.; Zemb, T.; Dubois, M.; Cölfen, H. Osmotic Pressure and Phase Boundary Determination of Multiphase Systems by Analytical Ultracentrifugation. *ChemPhysChem* **2008**, *9* (6), 882–890.

(28) van Rijssel, J.; Peters, V. F.; Meeldijk, J. D.; Kortschot, R. J.; van Dijk-Moes, R. J.; Petukhov, A. V.; Erne, B. H.; Philipse, A. P. Size-Dependent Second Virial Coefficients of Quantum Dots from Quantitative Cryogenic Electron Microscopy. *Journal of Physical Chemistry B* **2014**, *118* (37), 11000–11005. https://doi.org/10.1021/jp5056182.